\documentclass[10pt,a4paper,onecolumn]{article}
\usepackage{marginnote}
\usepackage{graphicx}
\usepackage{xcolor}
\usepackage{authblk,etoolbox}
\usepackage{titlesec}
\usepackage{calc}
\usepackage{tikz}
\usepackage{hyperref}
\hypersetup{colorlinks,breaklinks,
            urlcolor=[rgb]{0.0, 0.5, 1.0},
            linkcolor=[rgb]{0.0, 0.5, 1.0}}
\usepackage{caption}
\usepackage{tcolorbox}
\usepackage{amssymb,amsmath}
\usepackage{ifxetex,ifluatex}
\usepackage{seqsplit}
\usepackage{xstring}

\usepackage{fixltx2e} 
\usepackage[
  backend=biber,
]{biblatex}
\bibliography{paper.bib}

\let\textttOrig=\texttt
\def\texttt#1{\expandafter\textttOrig{\seqsplit{#1}}}
\renewcommand{\seqinsert}{\ifmmode
  \allowbreak
  \else\penalty6000\hspace{0pt plus 0.02em}\fi}

\makeatletter
\let\href@Orig=\href
\def\href@Urllike#1#2{\href@Orig{#1}{\begingroup
    \def\Url@String{#2}\Url@FormatString
    \endgroup}}
\def\href@Notdoi#1#2{\def\tempa{#1}\def\tempb{#2}%
  \ifx\tempa\tempb\relax\href@Urllike{#1}{#2}\else
  \href@Orig{#1}{#2}\fi}
\def\href#1#2{%
  \IfBeginWith{#1}{https://doi.org}%
  {\href@Urllike{#1}{#2}}{\href@Notdoi{#1}{#2}}}
\makeatother

\usepackage[top=3.5cm, bottom=3cm, right=1.5cm, left=1.0cm,
            headheight=2.2cm, reversemp, includemp, marginparwidth=4.5cm]{geometry}

\titleformat{\section}
  {\normalfont\sffamily\Large\bfseries}
  {}{0pt}{}
\titleformat{\subsection}
  {\normalfont\sffamily\large\bfseries}
  {}{0pt}{}
\titleformat{\subsubsection}
  {\normalfont\sffamily\bfseries}
  {}{0pt}{}
\titleformat*{\paragraph}
  {\sffamily\normalsize}

\usepackage{fancyhdr}
\makeatletter
\let\ps@plain\ps@fancy
\fancyheadoffset[L]{4.5cm}
\fancyfootoffset[L]{4.5cm}

\definecolor{linky}{rgb}{0.0, 0.5, 1.0}

\newtcolorbox{repobox}
   {colback=red, colframe=red!75!black,
     boxrule=0.5pt, arc=2pt, left=6pt, right=6pt, top=3pt, bottom=3pt}

\newcommand{\ExternalLink}{%
   \tikz[x=1.2ex, y=1.2ex, baseline=-0.05ex]{%
       \begin{scope}[x=1ex, y=1ex]
           \clip (-0.1,-0.1)
               --++ (-0, 1.2)
               --++ (0.6, 0)
               --++ (0, -0.6)
               --++ (0.6, 0)
               --++ (0, -1);
           \path[draw,
               line width = 0.5,
               rounded corners=0.5]
               (0,0) rectangle (1,1);
       \end{scope}
       \path[draw, line width = 0.5] (0.5, 0.5)
           -- (1, 1);
       \path[draw, line width = 0.5] (0.6, 1)
           -- (1, 1) -- (1, 0.6);
       }
   }

\patchcmd{\@maketitle}{center}{flushleft}{}{}
\patchcmd{\@maketitle}{center}{flushleft}{}{}
\patchcmd{\@maketitle}{\LARGE}{\LARGE\sffamily}{}{}
\def\maketitle{{%
  
  \AB@maketitle}}
\makeatletter
\renewcommand\AB@affilsepx{ \protect\Affilfont}
\renewcommand\AB@affilnote[1]{{\bfseries #1}\hspace{3pt}}
\renewcommand{\affil}[2][]%
   {\newaffiltrue\let\AB@blk@and\AB@pand
      \if\relax#1\relax\def\AB@note{\AB@thenote}\else\def\AB@note{#1}%
        \setcounter{Maxaffil}{0}\fi
        \begingroup
        \let\href=\href@Orig
        \let\texttt=\textttOrig
        \let\protect\@unexpandable@protect
        \def\thanks{\protect\thanks}\def\footnote{\protect\footnote}%
        \@temptokena=\expandafter{\AB@authors}%
        {\def\\{\protect\\\protect\Affilfont}\xdef\AB@temp{#2}}%
         \xdef\AB@authors{\the\@temptokena\AB@las\AB@au@str
         \protect\\[\affilsep]\protect\Affilfont\AB@temp}%
         \gdef\AB@las{}\gdef\AB@au@str{}%
        {\def\\{, \ignorespaces}\xdef\AB@temp{#2}}%
        \@temptokena=\expandafter{\AB@affillist}%
        \xdef\AB@affillist{\the\@temptokena \AB@affilsep
          \AB@affilnote{\AB@note}\protect\Affilfont\AB@temp}%
      \endgroup
       \let\AB@affilsep\AB@affilsepx
}
\makeatother

\renewcommand\Affilfont{\sffamily\small\mdseries}
\ifnum 0\ifxetex 1\fi\ifluatex 1\fi=0 
  \usepackage[T1]{fontenc}
  \usepackage[utf8]{inputenc}

\else 
  \ifxetex
    \usepackage{mathspec}
  \else
    \usepackage{fontspec}
  \fi
  \defaultfontfeatures{Ligatures=TeX,Scale=MatchLowercase}

\fi
\IfFileExists{upquote.sty}{\usepackage{upquote}}{}
\IfFileExists{microtype.sty}{%
\usepackage{microtype}
\UseMicrotypeSet[protrusion]{basicmath} 
}{}

\usepackage{hyperref}
\hypersetup{unicode=true,
            pdftitle={anesthetic: nested sampling visualisation},
            pdfborder={0 0 0},
            breaklinks=true}
\let\addcontentslineOrig=\addcontentsline
\def\addcontentsline#1#2#3{\bgroup
  \let\texttt=\textttOrig\addcontentslineOrig{#1}{#2}{#3}\egroup}
\let\markbothOrig\markboth
\def\markboth#1#2{\bgroup
  \let\texttt=\textttOrig\markbothOrig{#1}{#2}\egroup}
\let\markrightOrig\markright
\def\markright#1{\bgroup
  \let\texttt=\textttOrig\markrightOrig{#1}\egroup}

\usepackage{graphicx,grffile}
\makeatletter
\def\maxwidth{\ifdim\Gin@nat@width>\linewidth\linewidth\else\Gin@nat@width\fi}
\def\maxheight{\ifdim\Gin@nat@height>\textheight\textheight\else\Gin@nat@height\fi}
\makeatother
\setkeys{Gin}{width=\maxwidth,height=\maxheight,keepaspectratio}
\IfFileExists{parskip.sty}{%
\usepackage{parskip}
}{
\setlength{\parindent}{0pt}
\setlength{\parskip}{6pt plus 2pt minus 1pt}
}
\providecommand{\tightlist}{%
  \setlength{\itemsep}{0pt}\setlength{\parskip}{0pt}}
\ifx\paragraph\undefined\else
\let\oldparagraph\paragraph
\renewcommand{\paragraph}[1]{\oldparagraph{#1}\mbox{}}
\fi
\ifx\subparagraph\undefined\else
\let\oldsubparagraph\subparagraph
\renewcommand{\subparagraph}[1]{\oldsubparagraph{#1}\mbox{}}
\fi

\title{anesthetic: nested sampling visualisation}

        \author[1, 2, 3]{Will Handley}
    
      \affil[1]{Astrophysics Group, Cavendish Laboratory, J.J.Thomson Avenue, Cambridge,
CB3 0HE, UK}
      \affil[2]{Kavli Institute for Cosmology, Madingley Road, Cambridge, CB3 0HA, UK}
      \affil[3]{Gonville \& Caius College, Trinity Street, Cambridge, CB2 1TA, UK}
  \date{\vspace{-5ex}}

\begin{document}
\maketitle

\marginpar{
  \sffamily\small

  {\bfseries DOI:} \href{https://doi.org/10.21105/joss.00538}{\color{linky}{10.21105/joss.00538}}

  \vspace{2mm}

  {\bfseries Software}
  \begin{itemize}
    \setlength\itemsep{0em}
    \item \href{https://github.com/openjournals/joss-reviews/issues/1414}{\color{linky}{Review}} \ExternalLink
    \item \href{https://github.com/williamjameshandley/anesthetic}{\color{linky}{Repository}} \ExternalLink
    \item \href{10.5281/zenodo.2614017}{\color{linky}{Archive}} \ExternalLink
  \end{itemize}

  \vspace{2mm}

  {\bfseries Submitted:} 22 April 2019\\
  {\bfseries Published:} 12 May 2019

  \vspace{2mm}
  {\bfseries License}\\
  Authors of papers retain copyright and release the work under a Creative Commons Attribution 4.0 International License (\href{http://creativecommons.org/licenses/by/4.0/}{\color{linky}{CC-BY}}).
}

\hypertarget{summary}{%
\section{Summary}\label{summary}}

\texttt{anesthetic} is a Python package for processing nested sampling
runs, and will be useful for any scientist or statistician who uses
nested sampling software. \texttt{anesthetic} unifies many existing
tools and techniques in an extensible framework that is intuitive for
users familiar with the standard Python packages, namely
\href{https://www.numpy.org/}{NumPy},
\href{https://www.scipy.org/}{SciPy},
\href{https://matplotlib.org/}{Matplotlib} and
\href{https://pandas.pydata.org/}{pandas}. It has been extensively used
in recent cosmological papers (W. Handley and Lemos 2019a, 2019b).

\hypertarget{nested-sampling}{%
\section{Nested sampling}\label{nested-sampling}}

Nested sampling (Skilling 2006) is an alternative to
Markov-Chain-Monte-Carlo techniques (Hastings 1970). Given some data
\(D\), for a scientific model \(M\) with free parameters \(\theta\),
Bayes theorem states:

\[ P(\theta|D) = \frac{P(D|\theta) P(\theta)}{P(D)}. \]

Traditional MCMC approaches ignore the Bayesian evidence \(P(D)\) and
instead focus on the problem of generating samples from the posterior
\(P(\theta|D)\) using knowledge of the prior \(P(\theta)\) and
likelihood \(P(D|\theta)\). Nested sampling reverses this priority, and
instead computes the evidence \(P(D)\) (the critical quantity in
Bayesian model comparison (Trotta 2008)), producing posterior samples as
a by-product. Nested sampling does this by evolving a set of live points
drawn from the prior under a hard likelihood constraint which steadily
increases, causing the live points to contract around the peak(s) of the
likelihood. The history of the live-point evolution can be used to
reconstruct both the evidence and posterior samples, as well as the
density of states and consequently the full partition function.

Current publicly available implementations of nested sampling include
MultiNest (Feroz, Hobson, and Bridges 2009), PolyChord (W. J. Handley,
Hobson, and Lasenby 2015a, 2015b; Higson 2018), DNest (Brewer and
Foreman-Mackey 2018) and dynesty (Speagle 2019), all of which have been
incorporated into a wide range of cosmological (Lewis and Bridle 2002;
Zuntz et al. 2015; Brinckmann and Lesgourgues 2019) and particle physics
(The GAMBIT Scanner Workgroup: 2017) codes.

\begin{figure}
\centering
\includegraphics{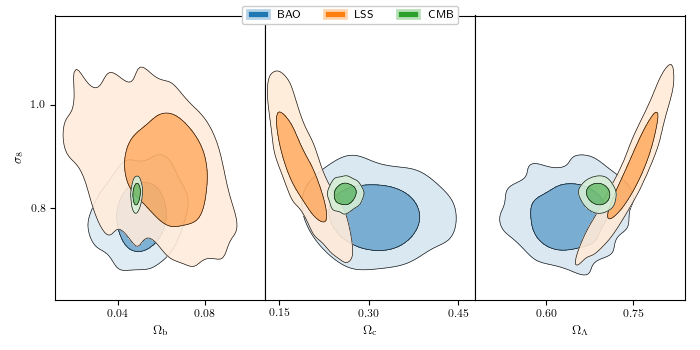}
\caption{Marginalised posterior plots produced by \texttt{anesthetic}.
The x axes indicate the fraction of normal matter, dark matter and dark
energy respectively, whilst the y-axis is the amplitude of mass
fluctuation in our late-time universe. The three measurements were
performed using measurements of baryonic acoustic oscillations, large
scale structure and the cosmic microwave background (W. Handley and
Lemos 2019a). It is an open cosmological and statistical questions
whether the LSS and CMB are consistent with one another.}
\end{figure}

\hypertarget{anesthetic}{%
\section{aNESThetic}\label{anesthetic}}

\texttt{anesthetic} acts on outputs of nested sampling software
packages. It can:

\begin{enumerate}
\def\labelenumi{\arabic{enumi}.}
\tightlist
\item
  Compute inferences of the Bayesian evidence (Trotta 2008), the
  Kullback-Leibler divergence (Kullback and Leibler 1951) of the
  distribution, the Bayesian model dimensionality (W. Handley and Lemos
  2019b) and the full partition function.
\item
  Dynamically replay nested sampling runs.
\item
  Produce one- and two-dimensional marginalised posterior plots (Figure
  1).
\end{enumerate}

A subset of computations from item 1 is provided by many of the nested
sampling software packages. \texttt{anesthetic} allows you to compute
these independently and more accurately, providing a unified set of
outputs and separating these computations from the generation of nested
samples.

Item 2 is useful for users that have experienced the phenomenon of `live
point watching' -- the process of continually examining the evolution of
the live points as the run progresses in an attempt to diagnose problems
in likelihood and/or sampling implementations. The GUI provided allows
users to fully reconstruct the run at any iteration, and examine the
effect of dynamically adjusting the thermodynamic temperature.

Finally, it is important to recognise that the functionality from item 3
is also provided by many other high-quality software packages, such as
getdist (Lewis 2015),
\href{https://corner.readthedocs.io/en/latest/}{corner} (Foreman-Mackey
2016), \href{https://pygtc.readthedocs.io/en/latest/}{pygtc} (Bocquet
and Carter 2016), \href{https://dynesty.readthedocs.io}{dynesty}
(Speagle 2019) and
\href{http://baudren.github.io/montepython.html}{MontePython}
(Brinckmann and Lesgourgues 2019). \texttt{anesthetic} adds to this
functionality by:

\begin{itemize}
\tightlist
\item
  Performing kernel density estimation using the state-of-the-art
  \href{https://pypi.org/project/fastkde/}{fastkde} (O'Brien et al.
  2016) algorithm.
\item
  Storing samples and plotting grids as a weighted
  \texttt{pandas.DataFrame}, which is more consistent with the
  scientific Python canon, allows for unambiguous access to samples and
  plots via their reference names, and easy definition of new
  parameters.
\item
  Using a contour colour scheme that is better suited to plotting
  distributions with uniform probability, which is important if one
  wishes to plot priors along with posteriors.
\end{itemize}

The source code for \texttt{anesthetic} is available on GitHub, with its
automatically generated documentation at
\href{https://anesthetic.readthedocs.io/}{ReadTheDocs} and a
pip-installable package on
\href{https://pypi.org/project/anesthetic/}{PyPi}. An example
interactive Jupyter notebook is given using
\href{https://mybinder.org/v2/gh/williamjameshandley/anesthetic/master?filepath=demo.ipynb}{Binder}
(Jupyter et al. 2018). Continuous integration is implemented with
\href{https://travis-ci.org/williamjameshandley/anesthetic}{Travis} and
\href{https://circleci.com/gh/williamjameshandley/anesthetic}{Circle}.

\hypertarget{acknowledgements}{%
\section{Acknowledgements}\label{acknowledgements}}

Bug-testing was provided by Pablo Lemos.

\hypertarget{references}{%
\section*{References}\label{references}}
\addcontentsline{toc}{section}{References}

\hypertarget{refs}{}
\leavevmode\hypertarget{ref-pygtc}{}%
Bocquet, Sebastian, and Faustin W. Carter. 2016. ``Pygtc: Beautiful
Parameter Covariance Plots (Aka. Giant Triangle Confusograms).''
\emph{The Journal of Open Source Software} 1 (6).
\url{https://doi.org/10.21105/joss.00046}.

\leavevmode\hypertarget{ref-dnest}{}%
Brewer, Brendon, and Daniel Foreman-Mackey. 2018. ``DNest4: Diffusive
Nested Sampling in C++ and Python.'' \emph{Journal of Statistical
Software, Articles} 86 (7): 1--33.
\url{https://doi.org/10.18637/jss.v086.i07}.

\leavevmode\hypertarget{ref-montepython}{}%
Brinckmann, Thejs, and Julien Lesgourgues. 2019. ``MontePython 3:
Boosted MCMC Sampler and Other Features.'' \emph{Physics of the Dark
Universe} 24: 100260. \url{https://doi.org/10.1016/j.dark.2018.100260}.

\leavevmode\hypertarget{ref-multinest}{}%
Feroz, F., M. P. Hobson, and M. Bridges. 2009. ``MULTINEST: an efficient
and robust Bayesian inference tool for cosmology and particle physics.''
\emph{Monthly Notices of the Royal Astronomical Society} 398 (October):
1601--14. \url{https://doi.org/10.1111/j.1365-2966.2009.14548.x}.

\leavevmode\hypertarget{ref-corner}{}%
Foreman-Mackey, Daniel. 2016. ``Corner.py: Scatterplot Matrices in
Python.'' \emph{The Journal of Open Source Software} 24.
\url{https://doi.org/10.21105/joss.00024}.

\leavevmode\hypertarget{ref-tension}{}%
Handley, Will, and Pablo Lemos. 2019a. ``Quantifying tension:
interpreting the DES evidence ratio.'' \emph{arXiv E-Prints}, February,
arXiv:1902.04029. \url{http://arxiv.org/abs/1902.04029}.

\leavevmode\hypertarget{ref-dimensionality}{}%
Handley, Will, and Pablo Lemos. 2019b. ``Quantifying dimensionality: Bayesian cosmological
model complexities.'' \emph{arXiv E-Prints}, March, arXiv:1903.06682.
\url{http://arxiv.org/abs/1903.06682}.

\leavevmode\hypertarget{ref-polychord0}{}%
Handley, W. J., M. P. Hobson, and A. N. Lasenby. 2015a. ``POLYCHORD:
nested sampling for cosmology.'' \emph{Monthly Notices of the Royal
Astronomical Society} 450 (June): L61--L65.
\url{https://doi.org/10.1093/mnrasl/slv047}.

\leavevmode\hypertarget{ref-polychord1}{}%
Handley, W. J., M. P. Hobson, and A. N. Lasenby. 2015b. ``POLYCHORD: next-generation nested sampling.''
\emph{Monthly Notices of the Royal Astronomical Society} 453 (November):
4384--98. \url{https://doi.org/10.1093/mnras/stv1911}.

\leavevmode\hypertarget{ref-mcmc}{}%
Hastings, W. K. 1970. ``Monte Carlo Sampling Methods Using Markov Chains
and Their Applications.'' \emph{Biometrika} 57 (1): 97--109.
\url{https://doi.org/10.2307/2334940}.

\leavevmode\hypertarget{ref-dypolychord}{}%
Higson, Edward. 2018. ``dyPolyChord: Dynamic Nested Sampling with
PolyChord.'' \emph{Journal of Open Source Software} 3 (29): 916.
\url{https://doi.org/10.21105/joss.00965}.

\leavevmode\hypertarget{ref-binder}{}%
Jupyter et al. 2018. ``Binder 2.0 - Reproducible, Interactive, Sharable
Environments for Science at Scale.'' In. Proceedings of the 17th Python
in Science Conference.
\url{https://doi.org/10.25080/Majora-4af1f417-011}.

\leavevmode\hypertarget{ref-KL}{}%
Kullback, S., and R. A. Leibler. 1951. ``On Information and
Sufficiency.'' \emph{Ann. Math. Statist.} 22 (1): 79--86.
\url{https://doi.org/10.1214/aoms/1177729694}.

\leavevmode\hypertarget{ref-getdist}{}%
Lewis, Anthony. 2015. ``Getdist Github Repository.''
\url{https://github.com/cmbant/getdist}.

\leavevmode\hypertarget{ref-cosmomc}{}%
Lewis, Antony, and Sarah Bridle. 2002. ``Cosmological parameters from
CMB and other data: A Monte Carlo approach.'' \emph{Phys. Rev.} D66:
103511. \url{https://doi.org/10.1103/PhysRevD.66.103511}.

\leavevmode\hypertarget{ref-fastkde}{}%
O'Brien, Travis A., Karthik Kashinath, Nicholas R. Cavanaugh, William D.
Collins, and John P. O'Brien. 2016. ``A Fast and Objective
Multidimensional Kernel Density Estimation Method: FastKDE.''
\emph{Computational Statistics \& Data Analysis} 101: 148--60.
\url{https://doi.org/10.1016/j.csda.2016.02.014}.

\leavevmode\hypertarget{ref-skilling}{}%
Skilling, John. 2006. ``Nested Sampling for General Bayesian
Computation.'' \emph{Bayesian Analysis.} 1 (4): 833--59.
\url{https://doi.org/10.1214/06-BA127}.

\leavevmode\hypertarget{ref-dynesty}{}%
Speagle, Joshua S. 2019. ``dynesty: A Dynamic Nested Sampling Package
for Estimating Bayesian Posteriors and Evidences.'' \emph{arXiv
E-Prints}, April, arXiv:1904.02180.
\url{http://arxiv.org/abs/1904.02180}.

\leavevmode\hypertarget{ref-gambit}{}%
The GAMBIT Scanner Workgroup:. 2017. ``Comparison of Statistical
Sampling Methods with Scannerbit, the Gambit Scanning Module.''
\emph{The European Physical Journal C} 77 (11): 761.
\url{https://doi.org/10.1140/epjc/s10052-017-5274-y}.

\leavevmode\hypertarget{ref-trotta}{}%
Trotta, R. 2008. ``Bayes in the sky: Bayesian inference and model
selection in cosmology.'' \emph{Contemporary Physics} 49 (March):
71--104. \url{https://doi.org/10.1080/00107510802066753}.

\leavevmode\hypertarget{ref-cosmosis}{}%
Zuntz, J., M. Paterno, E. Jennings, D. Rudd, A. Manzotti, S. Dodelson,
S. Bridle, S. Sehrish, and J. Kowalkowski. 2015. ``CosmoSIS: Modular
cosmological parameter estimation.'' \emph{Astronomy and Computing} 12
(September): 45--59. \url{https://doi.org/10.1016/j.ascom.2015.05.005}.

\end{document}